\documentclass{article}
\usepackage{spconf,amsmath,graphicx}
\usepackage{color}
\usepackage{amsmath,amsfonts,bm,cite,url}
\usepackage[activate]{microtype}
\usepackage{subfigure}
\usepackage[export]{adjustbox}

\newcommand{\setC}{\ensuremath{\mathbb{C}}}

\DeclareMathOperator*{\argmax}{arg\,max}

\title{Dual-Encoder Architecture with Encoder Selection for Joint Close-Talk and Far-Talk Speech Recognition}
\name{Felix Weninger$^1$, Marco Gaudesi$^2$, Ralf Leibold$^3$, Roberto Gemello$^2$, Puming Zhan$^1$}
\address{$^1$Nuance Communications, Burlington, MA, USA\\$^2$Nuance Communications, Torino, Italy\\$^3$Nuance Communications, Aachen, Germany\\{\small \tt \{felix.weninger,marco.gaudesi,ralf.leibold,roberto.gemello,puming.zhan\}@nuance.com}}
\begin{document}
\ninept
\maketitle
\begin{abstract}
In this paper, we propose a dual-encoder ASR architecture for joint modeling of close-talk (CT) and far-talk (FT) speech, in order to combine the advantages of CT and FT devices for better accuracy.
The key idea is to add an encoder selection network to choose the optimal input source (CT or FT) and  the corresponding encoder.
We use a single-channel encoder for CT speech and a multi-channel encoder with Spatial Filtering neural beamforming for FT speech, which are jointly trained with the encoder selection.
We validate our approach on both attention-based and RNN Transducer end-to-end ASR systems.
The experiments are done with conversational speech from a medical use case, which is recorded simultaneously with a CT device and a microphone array. 
Our results show that the proposed dual-encoder architecture obtains up to 9\% relative WER reduction when using both CT and FT input, compared to the best single-encoder system trained and tested in matched condition. 
\end{abstract}
\section{Introduction}
\label{sec:intro}



Improving the accuracy of far-field ASR remains a challenging problem, despite decades of research on the topic \cite{VanCompernolle1990-SRI,Yoshioka2012-MMU,Li2014-AOO,HaebUmbach2021-FAS}.
In recent years, the end-to-end (E2E) modeling technique  \cite{Graves2012-STW,Chan2016-LAA,Weninger2019-LAS,Zeyer2019-ACO,Zhang2020-TTA,Gulati2020-CCA,Tuske2020-SHA}, which replaces the components of a traditional ASR system by a single neural network, has been proven to be advantageous for recognition accuracy in general and for far-field ASR in particular.
In multi-channel E2E ASR, the neural network also includes neural beamforming  \cite{Sainath16-RTC,Erdogan2016-IMB,Li2017-AMF,Ochiai2017-UAF,Chang2020-EMS}, which replaces the traditional multi-channel frontend.
Despite these advances, there still exists a gap between the performance of ASR systems on close-talk (CT) and far-talk (FT) speech \cite{HaebUmbach2021-FAS}.

In general, CT devices, such as headset microphones, are best at capturing a single speaker who is wearing the device, but they are sometimes impractical or obtrusive to use.
On the contrary, FT devices, such as microphone arrays, are well suited for capturing multiple speakers; thus, they are popular for tasks such as meeting transcription \cite{Araki2018-MRW,Yoshioka2019-MTU,Wang2020-EEM}.
The motivation for our work is to combine the advantages of CT and FT devices by designing a single E2E ASR system that uses both CT and FT input simultaneously, in order to achieve better accuracy.
For this, it is not only important to automatically select the best input source for a given speech utterance, but also to train dedicated components of the ASR system for CT and FT speech: It is known that an ASR system trained with CT data performs much better on CT test data and significantly worse on FT test data, and vice versa \cite{Tang2018-ASO}.

Thus, the main contribution of this paper is to introduce a multi-channel, multi-encoder E2E ASR architecture where an encoder selection network selects the optimal input source (CT or FT) and passes it through the matching encoder.
The encoder selection network bases its decision on the input features, which capture important information such as signal quality of each input source and speaker role / identity.
Since the encoders are both connected to a common decoder, this architecture bears some similarity to a hybrid system with different acoustic models, one for CT and one for FT.
However, in our work, the whole system, including the encoder selection and the beamforming frontend, is trained end-to-end.

The proposed architecture with encoder selection has the following advantages:
First, assuming a well trained encoder selection network, each encoder is trained and tested with matching data.
This means that the system can be operated with only a single input source (CT or FT) and achieve similar performance as the corresponding single-encoder system in matched condition.
This is unlike multi-style or multi-condition training \cite{Lippmann1987-MST}, which often performs suboptimally compared to the matched condition scenario.
Second, the system can perform both `hard' and `soft' encoder selection (i.e.\ compute a weighted average of the encoders) in inference. 
The benefit of hard selection is that only one of the encoder needs to be evaluated for each utterance, thus keeping a similar computational cost as a single-encoder system.
Conversely, using `soft' encoder selection can further improve the performance (at increased computational cost), similar to traditional model combination approaches for hybrid systems such as \cite{Misra2003-NEB}.

\section{Related work}
\label{sec:related}

To our knowledge, \cite{Chen17-MTI} was the first to propose a `dual-encoder' sequence-to-sequence system in the area of natural language understanding, yet without encoder selection as in our work.

In the `bifocal' ASR approach \cite{Macoskey2021-BNA}, two encoder networks are used to reduce latency for keyword spotting by using a small encoder for wake word detection and a large one for ASR. Unlike our method, both encoders use the same input and are used in mutually exclusive execution.

In a similar vein, \cite{Narayanan2021-CEF} uses two encoders in a single E2E model in order to unify streaming and non-streaming ASR. However, unlike in our work, there is only a single type of input to the system, and the encoders are cascaded, not run in parallel.

Multi-channel attention \cite{Braun2018-MCA,Ochiai2017-UAF,Chang2021-EMT} can be understood as selecting an optimal input source by an attention layer. The key difference to our approach is that the same acoustic model is used regardless of which input channel is preferred.


In the `multi-stream' E2E ASR approach \cite{Wang2019-SAB,Li2020-MSE,Li2021-TAA}, a hierarchical attention network is used to focus the decoder first on one stream (i.e.\ encoder) and then on a certain position within the encoder output.
Our work differs from multi-stream E2E ASR in the following aspects:
First, by using an encoder selection network based on the input features instead of an attention mechanism based on the encoder outputs, we can skip all but one of the encoders in inference, thus keeping the inference cost similar to the single-encoder architecture
(note that in \cite{Li20-APT}, a two-stage strategy was proposed to improve training efficiency, but without reducing inference time).
Moreover, the aforementioned body of work is limited to the attention-based E2E architecture by design, whereas our approach also works for other types of E2E architectures (in particular, the RNN-T architecture),
and it does not consider E2E training including the multi-channel frontend (i.e.\ neural beamforming) as in our paper.




\section{End-to-end ASR architectures}

In this section, we briefly describe the basic E2E ASR architectures used for our work before introducing our multi-encoder approach.

\subsection{Encoder-decoder architecture with attention}

First, we use an encoder-decoder architecture with attention similar to Listen-Attend-Spell (LAS)\cite{Chan2016-LAA,Weninger2019-LAS}.
The encoder $e$ generates a hidden representation $e(x)$ of the input features $x$.
In our work, $e$ is implemented as a stack of bidirectional Long Short-Term Memory (bLSTM) layers. 
The decoder determines the distribution $p_i = p(y_i | y_1, \dots, y_{i-1}, x)$ for the $i$-th symbol in the output sequence.
It implicitly aligns the encoder outputs with the output sequence by computing the context vector $c_i$ as weighted sum of $e(x)$, using the Bahdanau attention mechanism \cite{Bahdanau2015-NMT}.

\subsection{RNN Transducer}

Second, we validate our approach on the Recurrent Neural Network Transducer (RNN-T) architecture \cite{Graves2012-STW,Zhang2020-TTA}.
The RNN-T consists of an encoder, a prediction network and a joint network.
The prediction network is similar to an RNN language model which computes the distribution $p(y_i|y_1,\dots,y_{i-1})$.
The joint network is a feed-forward network that computes the alignment probability $p(z_i|x,t_i,y_1,\dots,y_{i-1})$ of the output symbol $z_i$ with the encoder output $e(x)$ at time frame $t_i$.
The RNN-T approach computes the probability $p(y|x)$ of the output sequence given the input by marginalizing over the possible alignments $z$.
In our paper, the encoder $e$ consists of Conformer \cite{Gulati2020-CCA} layers.

\subsection{Multi-Encoder Architecture with Encoder Selection}

\begin{figure}[t]
\centering
\includegraphics[width=.85\columnwidth]{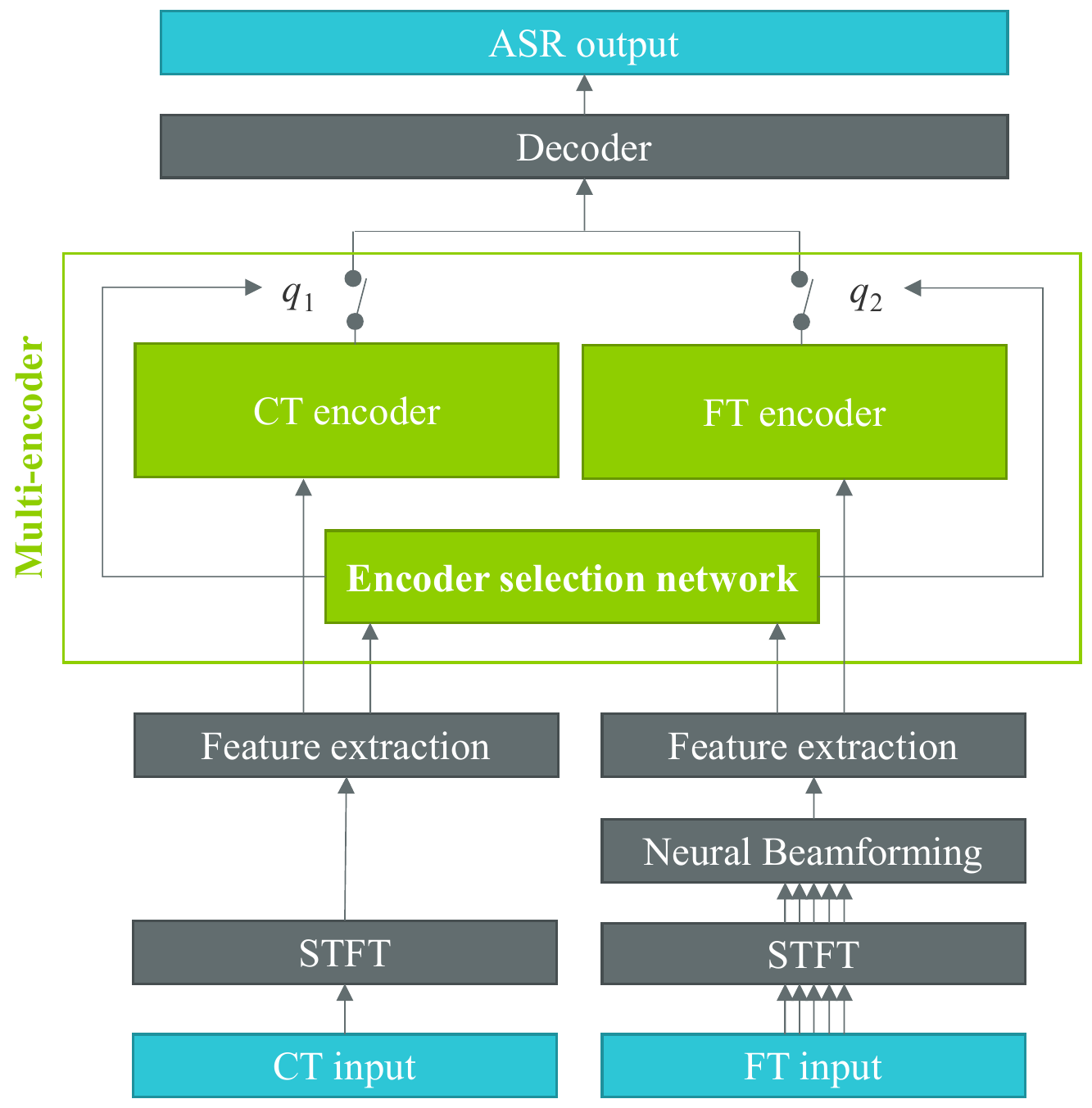}
\caption{Dual-encoder end-to-end ASR system with encoder selection network for joint modeling of single-channel close-talk (CT) and multi-channel far-talk (FT) input. 
The switch symbols indicate hard or soft encoder selection with selection probabilities $q_1, q_2$.
}
\label{fig:arch}
\end{figure}

In the multi-encoder architecture, the encoder $e$ consists of several sub-encoders $e^{(1)}, e^{(2)}, \dots, e^{(E)}$, which are combined using an encoder selection network as detailed below.
The architecture of the proposed multi-encoder system for CT and FT input ($E=2$) is shown in \figurename~\ref{fig:arch}.
Both encoders start from the raw waveform and the feature extraction is done on-the-fly.
In case of FT input, the features are extracted from a single channel signal obtained by neural beamforming.

\subsubsection{Encoder selection}

The encoder selection network computes the probability $q_k$ that the $k$-th sub-encoder is optimally suited for the current utterance. It takes as input the concatenation of the input features $x^{(1)}, x^{(2)}, \dots, x^{(E)}$ of the $E$ sub-encoders:
\begin{equation}
q_k = \text{EncoderSelection}( [ x^{(1)} ; \cdots ; x^{(E)} ] ).
\end{equation}

The topology of the encoder selection network corresponds to the sequence classification architecture proposed in \cite{Norouzian2019-EAM}.
It consists of convolutional and LSTM layers, followed by an attention layer \cite{Neumann2017-ACN} to perform summarization across an utterance and a softmax layer.

Using the encoder selection network as a classifier, we can decide to evaluate only one sub-encoder ({\em hard selection}):
\begin{equation}
e(x^{(1)}, \dots, x^{(E)}) = e^{(k^*)}(x^{(k^*)}), \quad k^* = \argmax_k q_k,
\label{eq:hard_sel}
\end{equation}
where $e^{(k)}$ is the hidden representation of the input features $x^{(k)}$ computed by the $k$-th sub-encoder.
Alternatively, the scores of the encoder selection network can be used to perform {\em soft selection}, i.e.\ compute a weighted average:
\begin{equation}
e(x^{(1)}, \dots, x^{(E)}) = \sum_k q_k e^{(k)}(x^{(k)}).
\label{eq:soft_sel}
\end{equation}
The result of the encoder selection (Eq.~\eqref{eq:hard_sel} or \eqref{eq:soft_sel}) is then processed by the decoder (or joint network in case of RNN-T) of the ASR system as usual.
The encoder selection network is trained jointly with the rest of the system.
During training, soft selection is always used, in order to keep the training objective differentiable.

We also consider frame-wise encoder selection and combination, where we remove the attention layer from the encoder selection network and replace it with an average pooling layer.
The pooling size is chosen in a way that the total frame decimation done in the encoder selection network (by convolution strides and average pooling) matches the decimation of the input by the sub-encoders.

\subsubsection{Far-talk encoder using neural beamforming}

In our dual-encoder system (\figurename~\ref{fig:arch}), a neural beamformer is used on the multi-channel input before extracting the FT features used in the FT encoder as well as the encoder selection network.
The neural beamformer is implemented via the Spatial Filtering (SF) layer \cite{Wu2019-FDM,Kumatani2019-MGS}, which is a complex-valued hidden layer designed to mimic a filter-and-sum beamforming operation with time-invariant beamforming coefficients.
The input is the multi-channel complex spectrum. 
The SF layer performs the following operation: 
\begin{equation}
    y_{t,f,d} = \sum_c w_{f,d,c} \, x_{t,f,c} + b_{f,d},
    \label{eq:sf}
\end{equation}
where $t$, $f$, $d$ and $c$ are the time frame, frequency, look direction and input channel indices, $w_{f,d,c} \in \setC$ is a trainable weight, $x_{t,f,c}$ is an STFT coefficient of the time-frequency bin $(t,f)$ of the $c$-th channel of the input signal, and $b_{f,d} \in \setC$ is a bias term.
After the SF layer, an average pooling layer is inserted to compute the enhanced power spectrogram $\hat{\bf X}$ : 
\begin{equation}
    \hat{x}_{t,f} = \frac{1}{D} \sum_d | y_{t,f,d} |^2 ,
\end{equation}
where $D$ is the number of look directions. 
The weights of the neural beamformer are trained jointly with the rest of the system, propagating the gradients through the FT feature extraction. 

\section{Experiments and Results}

\subsection{Experimental setup}

\subsubsection{Data sets}

Our experiments are carried out in a domain adaptation scenario.
A seed model with only a CT encoder is trained on a large corpus of doctor-patient conversations recorded with a CT device.
The seed model is then adapted to the FT input or joint CT / FT input.
An adaptation set of 40\,h of parallel CT / FT data was recorded in a similar medical conversation scenario. 
Only the doctor uses a CT device, while all speakers (doctor and others) are recorded by a wall-mounted 16-channel linear microphone array.
The CT and FT data are synchronized using a cross-correlation based method.
The evaluation is done on a test set that matches the adaptation set in terms of recording conditions, but contains data from different speakers. 
All data are anonymized.
For most of our experiments, we use manually end-pointed utterances.
The audio is downsampled to 16\,kHz sampling rate.

\subsubsection{Training of FT and joint CT / FT systems}

To obtain the single-encoder FT system, we clone the seed model, add the neural SF layer at the bottom of the encoder, and then fine-tune the model on the FT adaptation data using a small learning rate.
Similar to \cite{Wu2019-FDM}, the weights of the SF layer are initialized with pre-computed finite impulse response (FIR) filters designed with an ideal steering vector and an isotropic noise model.

To obtain the dual-encoder systems, we use the CT encoder from the seed model as initialization for both the CT and FT encoder. 
We add the encoder selection network, then retrain the resulting model end-to-end on the adaptation data, i.e.,
the encoder selection network is trained along with the other parts of the end-to-end model to maximize ASR accuracy (an alternative approach is presented in Section \ref{sec:pretrain}).
The SF layer of the FT encoder is initialized in the same way as for the single-encoder system.

\subsection{Experiments with attention-based system}

We first performed a set of experiments using the attention-based LAS architecture.
The topology is similar to the experiments we reported in \cite{Weninger2020-SSL}.
The acoustic features are 80-dimensional log-Mel features with a frame shift of 10\,ms and a window size of 32\,ms, which are batch normalized and cepstral mean normalized per utterance.
The encoders are composed of 6 bLSTM layers (size 512 for each direction) with an input decimation factor (stride) of 2 after every other layer, for a total stride of 8. 
The decoder consists of a single LSTM layer with size 1024. 
For the FT model, a SF layer with $D=11$ look directions and $C=16$ input channels is used.
Training of the seed model is done on 470 hours of CT data, using a training recipe similar to \cite{Weninger2020-SSL}.

SpecAugment \cite{Park2019-SAAa} is applied with $T_\text{max} = 50$, $m_T=2$, $F_\text{max}=30$, $m_F=1$.
The same SpecAugment mask is applied to the input features passed to each encoder.

In the dual-encoder system, the encoder selection network consists of 2 time-delay neural network layers with 256 units, 1 LSTM layer with 256 units, followed by an optional attention layer (for utterance-wise classification) and a softmax layer.
The LAS architecture with a single encoder has 58 M parameters, while the dual-encoder version has 87 M parameters (of which 1.1 M are used for the encoder selection network).

For comparison purpose, we also trained a `large' single-encoder model with a similar number of parameters (87 M).
This was done by increasing the width of the layers (size 670 per direction for the bLSTM encoder layers and size 1340 for the LSTM decoder layer).
Due to the enlarged size, we found it helpful to increase the dropout rate from 0.3 to 0.4 for training this model.

\begin{table}[t]
\centering
\begin{tabular}{l|l|l}
WER [\%] & CT input & FT input \\
\hline
\multicolumn{3}{c}{\em Single-encoder} \\
\hline
CT enc & 15.8 (12.3/23.8) & 18.9 (17.2/22.9) \\
SF(FT) enc & 16.4 (13.0/24.0) & 16.0 (14.3/19.7) \\
\hline
\multicolumn{3}{c}{\em Dual-encoder} \\
\hline
CT enc + SF(FT) enc & 16.0 (12.8/23.4) & 16.3 (14.7/19.7) \\
\end{tabular}
\caption{WER achieved by the LAS architecture on either close-talk (CT) or far-talk (FT) input.
The CT encoder uses the CT input or a single channel of the FT input.
The FT encoder uses Spatial Filtering (SF) beamforming for the FT input and skips the SF layer for CT input. WERs are given as Overall (Doctor/Others).}
\label{tab:single_input}
\end{table}

\subsubsection{Results on single input source (CT or FT)} 
\tablename~\ref{tab:single_input} shows the WER achieved by the LAS architecture on close-talk (CT) or far-talk (FT) input individually.
The LAS seed model trained on CT data obtains 15.8\,\% WER overall on CT and 18.9\,\% on FT input (using a single channel of the microphone array).
Comparing the results on FT input vs.\ CT input, we observe a large degradation for the doctor (12.3 to 17.2\,\%), while the other speakers are improved (23.8 to 22.9\,\%).
This is expected due to the distances of the speakers from the input devices, because the doctor is close to the CT device and the other speakers are closer to the FT device.

After finetuning the LAS model on the FT data using neural SF, the performance on the FT input is greatly enhanced (18.9 to 16.0\,\%).
The overall WER is similar to the CT input (15.8\,\%). 
On the one hand, there is large improvement for the other speakers (23.8 to 19.7\,\%); on the other hand, the WER on the doctor's speech is still far behind the CT case (14.3 vs.\ 12.3\,\%).
Moreover, the WER on the CT input (when removing the SF layer) is degraded from 15.8 to 16.4\,\%.

In contrast, the dual-encoder architecture achieves similar performance on CT and FT data (when using only one encoder) as the separate single-encoder systems which are trained and tested in matched condition. 
This is notable since the dual-encoder architecture shares the attention layer and decoder between CT and FT input, indicating that a single parameterization of attention and decoder can be effective for both types of inputs.

\begin{table}[t]
\centering
\begin{tabular}{l|c|c}
WER [\%] & Sel.~unit & CT + FT input  \\
\hline
\multicolumn{3}{c}{\em Single-encoder} \\
\hline
SF (CT; FT) enc & -- & 15.0 (12.9/19.6) \\
SF (CT; FT) enc (large) & -- & 15.5 (13.4/20.0) \\
\hline
\multicolumn{3}{c}{\em Dual-encoder} \\
\hline
CT + FT enc.\ hard sel. & Utt & 15.1 (13.0/19.8) \\
CT + FT enc.\ soft sel. & Utt & 14.4 (12.3/{\bf 19.2}) \\
CT + FT enc.\ hard sel. & Frame & 15.1 (13.0/19.7) \\
CT + FT enc.\ soft sel. & Frame & \bf 14.3 ({\bf 12.2}/{\bf 19.2}) \\
\hline
\end{tabular}
\caption{WER achieved by the LAS architecture on joint close-talk (CT) and far-talk (FT) input (reference segmentation). The single-encoder baseline uses Spatial Filtering (SF) beamforming on the channel-wise concatenation (CT; FT) of CT and FT input. The dual-encoder approach uses hard or soft encoder selection, and the encoder selection network is evaluated per utterance or per frame.}
\label{tab:dual_refseg}
\end{table}

\subsubsection{Results on combined CT and FT input}
\tablename~\ref{tab:dual_refseg} shows the performance achieved on combined CT and FT input.
As a baseline, we use a single-encoder system that uses neural SF applied to the channel-wise concatenation of the CT and FT input, denoted by SF (CT; FT).
This obtains a WER of 15.0\,\%, which is a sizeable improvement over the best result on single input (15.8\,\% for CT and 16.0\,\% on FT).
The dual-encoder architecture using utterance-based, hard encoder selection yields 15.1\,\% WER, which is similar to the SF baseline. 
However, further gains can be obtained by using soft selection, which yields 14.4\,\% WER overall (12.3\,\% for the doctor and 19.2\,\% for others).
This result is even slightly better than an oracle system combination of the CT and the FT single-encoder systems, where we assume the speaker role to be known and select the CT system for the doctor (resulting in 12.3\,\% WER) and the FT encoder system for the others (19.7\,\% WER).

Moreover, the dual-encoder architecture performs much better than the `large' single-encoder system with a similar number of parameters, which obtains 15.5\,\% WER. 
The reason that the large size system performs worse than the regular sized one is probably overfitting.
Finally, we also investigated the frame-wise encoder selection and combination and obtained similar results as with the utterance-wise encoder selection and combination.
This indicates that the optimal encoder remains the same throughout each test utterance.

\subsubsection{Results for VAD segmentation}
\tablename~\ref{tab:dual_vadseg} shows the results when we use speech segments obtained by voice activity detection (VAD) instead of manually segmented utterances.
Compared to the manual segmentation results, the absolute WERs are higher,
which can be attributed to the difficulty of reliable VAD for far-talk input \cite{HaebUmbach2021-FAS}.
However, the relative differences between the different architectures are similar to the test using manual segmentation, and the dual-encoder architecture retains its effectiveness.
For hard selection, the framewise encoder selection outperforms the utterance-wise selection, intuitively because more than one speaker can be active in a segment detected by VAD, and hence the optimal encoder (CT or FT) varies within the segment.
For soft selection however, the framewise and utterance-wise selection perform similar, indicating the robustness of the soft selection approach.
The reason why utterance-wise soft selection works reasonably well in the case of speaker switch might be its model averaging effect. For instance, the encoder selection network is likely to return a probability close to 0.5 if two speakers are equally present.

\begin{table}[t]
\centering
\begin{tabular}{l|c|c}
WER [\%] & Sel.~unit & CT + FT input  \\
\hline
\multicolumn{3}{c}{\em Single-encoder} \\
\hline
SF (CT; FT) & -- & 15.8 (13.6/20.6) \\
\hline
\multicolumn{3}{c}{\em Dual-encoder} \\
\hline
CT + FT enc.\ hard sel. & Utt & 16.4 (14.2/21.3) \\
CT + FT enc.\ soft sel. & Utt & {\bf 15.2} ({\bf 12.8}/20.4) \\
CT + FT enc.\ hard sel. & Frame & 16.0 (13.8/21.1) \\
CT + FT enc.\ soft sel. & Frame & {\bf 15.2} (13.0/{\bf 20.1}) \\
\hline
\end{tabular}
\caption{WER achieved by the LAS architecture on joint CT and FT input when using VAD to segment the input.}
\label{tab:dual_vadseg}
\end{table}

\begin{table*}[t]
\centering
\begin{tabular}{l|l|l|l}
WER [\%] & CT input & FT input & CT + FT input \\
\hline
\multicolumn{4}{c}{\em Single-encoder} \\
\hline
CT enc & 12.4 (9.6/18.5) & 14.0 (12.2/18.1) & -- \\
SF (FT) enc & 12.4 (9.7/18.5) $^1$ & 12.3 (10.6/16.3) & -- \\
SF (CT; FT) enc & 12.5 (9.7/18.6) $^1$ & 12.5 (10.7/16.3) $^2$ & 11.9 (10.1/16.0) \\
\hline
\multicolumn{4}{c}{\em Dual-encoder} \\
\hline
CT enc + SF (FT) enc & 12.6 (9.9/18.8) & 12.6 (10.9/16.4) & {\bf 11.6} ({\bf 9.6}/{\bf 16.0}) \\
\end{tabular}
\caption{WER achieved by the Conformer Transducer on close-talk (CT) and far-talk (FT) input, and on joint CT and FT input. The dual-encoder architecture uses utterance-wise soft encoder selection. $^1$: SF layer skipped for CT input; $^2$: CT input channel set to zero in SF}
\vspace{-2mm}
\label{tab:rnnt}
\end{table*}

\subsection{Experiments with Conformer Transducer}

To confirm the effectiveness of the multi-encoder approach, we performed additional experiments with the Conformer Transducer, using a topology similar to the `large' one from \cite{Gulati2020-CCA}.
The model uses the same acoustic features as the LAS architecture above.
The encoders have 16 Conformer layers of size 512 (feed-forward layer size 2\,048, convolution width 17).
The prediction network consists of a single LSTM layer of size 640, while the joint network has 512 units.
The single-encoder model has 104 M parameters in total.
Training is done on 3.8\,k hours of CT data, using a training recipe similar to \cite{Gulati2020-CCA}.

\tablename~\ref{tab:rnnt} shows the results obtained with the single- and dual-encoder Conformer Transducer.
The single-encoder CT seed model achieves 12.4\,\% and 14.0\,\% WER on CT and FT input, respectively.
The large drop in WER compared to the bLSTM LAS above is due to the improved architecture and enlarged training data.
Despite the difference in the absolute WER numbers, we generally observe the same trends as displayed in \tablename~\ref{tab:single_input} and \ref{tab:dual_refseg}.
Using a SF neural beamformer, 12.3\,\% WER are obtained on FT input, and 11.9\,\% WER on joint CT + FT input.
Unlike in the LAS case, the FT encoder also works well on the CT input when the SF layer is skipped.
This is probably due to the increased amount of training data used for the seed model, which makes the encoder more robust against different kinds of inputs.
We also evaluated the SF (CT; FT) approach on the FT input only, by setting the CT channel to zero. 
This performed well, yielding similar WER as the SF (FT) encoder on the FT input.

The dual-encoder Conformer Transducer performs slightly worse than the single-encoder baselines on the individual CT or FT inputs, but it obtains the overall best result on joint CT + FT input, i.e.\ 11.6\,\% WER.
The improvement compared to SF (CT; FT) is significant ($p<.001$) according to a matched pairs sentence-segment word error test \cite{Gillick1989-SSI}.
As in the LAS case, the dual-encoder with soft selection comes close to the oracle result when using the CT encoder for the doctor and the FT encoder for the others.

\subsubsection{Encoder selection pretraining}

\begin{figure}[t]
    \centering
    \subfigure[w/o pretraining]{
    \includegraphics[width=.8\columnwidth]{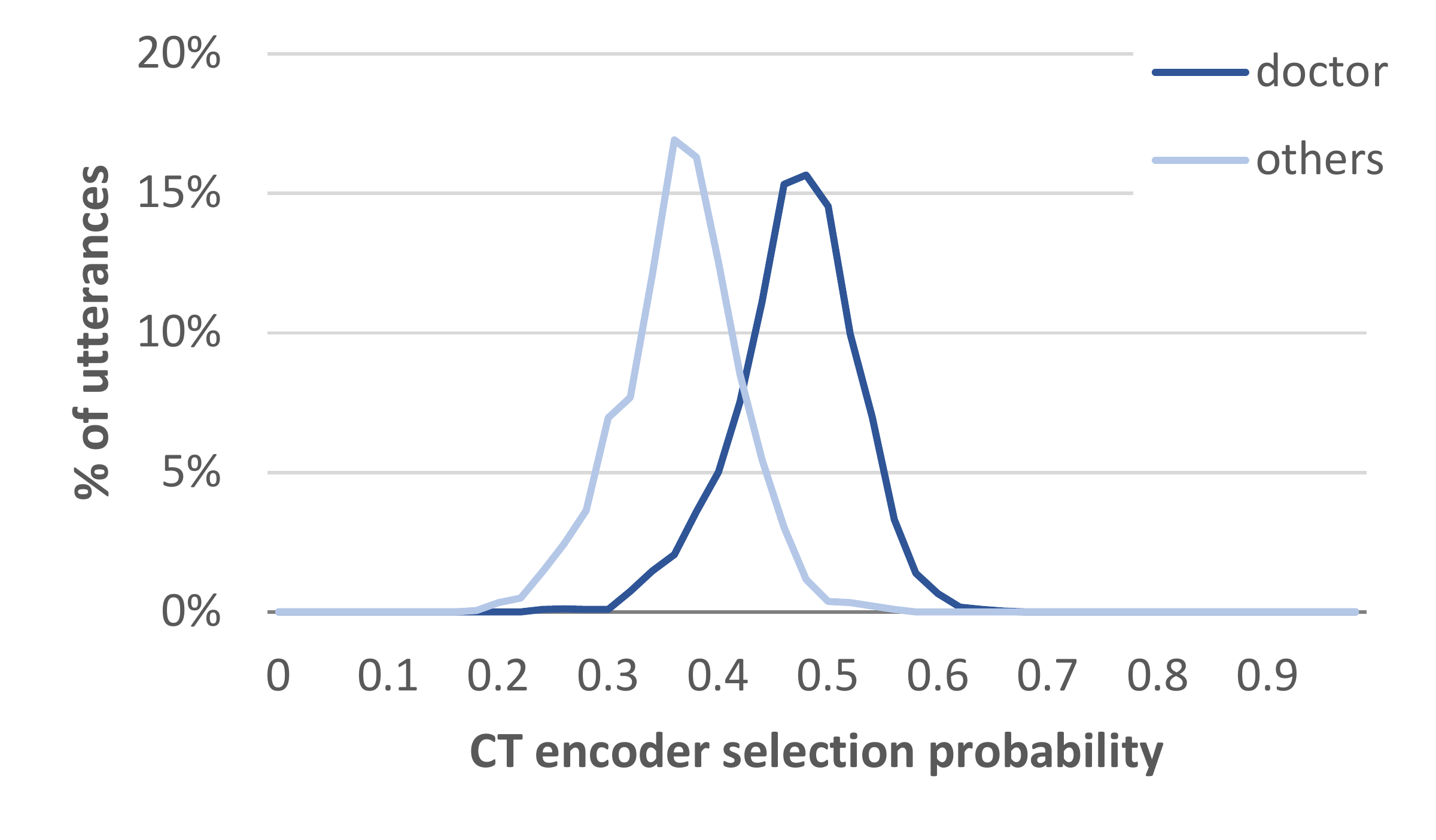}
    }
    \subfigure[w/ pretraining]{
    \includegraphics[width=.8\columnwidth]{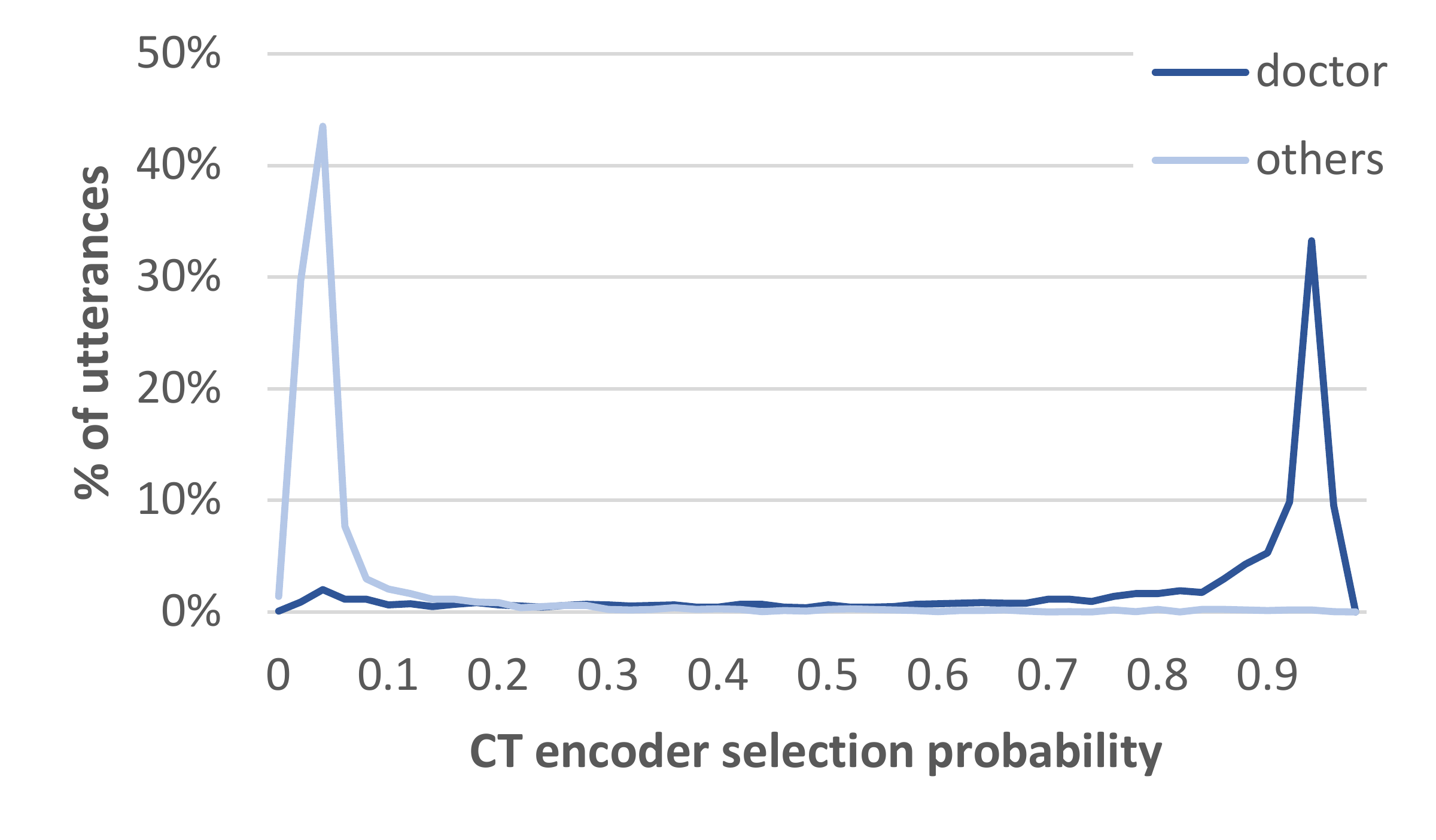}
    }
    \caption{Histogram of the CT encoder selection probability for doctor / other utterances in the test set, using the dual-encoder Conformer Transducer. Pretraining is done on the speaker role classification task (doctor / other).}
    \label{fig:histo}
\end{figure}

\label{sec:pretrain}

However, surprisingly we also found that {\em hard} encoder selection underperformed in case of the Conformer Transducer: the WER in this case was 12.4\,\%, which is similar to the single-encoder systems.
We examined the distribution of the encoder selection probability $q_1$ (i.e.\ the one for the CT encoder), which is shown as a histogram in \figurename~\ref{fig:histo}(a) for the utterances of the doctor vs.\ those of the other speakers in our test set.
It can be seen that the CT encoder selection probability for the doctor centers around 0.5, while it is lower for the other speakers (i.e.\ the FT encoder is preferred, as is expected).

Intuitively, we would expect the CT encoder to be selected with high probability for the doctor in our scenario.
This led us to investigate pretraining of the encoder selection network on the adaptation set with a sequence classification criterion, using the speaker role (doctor or other) as target for each utterance.
The pretrained encoder selection network is then kept frozen during the training of the dual-encoder model.
As a result, the distribution of the encoder selection probabilities changes as shown in \figurename~\ref{fig:histo}(b):
The confidence of the selection is largely increased, i.e.\ the encoder is selected with high probability for most of the utterances.
At the same time, we found that the WER of the system using hard encoder selection was improved significantly (12.4\,\% to 11.9\,\%, not shown in \tablename~\ref{tab:rnnt}).
This indicates that the speaker role classification is a good proxy task for the selection of the optimal encoder.
However, for soft selection, the pretraining did not yield any improvement. 
This means that while the joint training of the encoder selection network with the ASR system and the two-stage training lead to different solutions, they are both optimal in terms of the ASR performance with soft encoder selection -- which is the training criterion.


\subsubsection{Input synchronization and shift-aware training}

\begin{figure}
    \centering
    \includegraphics[width=.9\columnwidth]{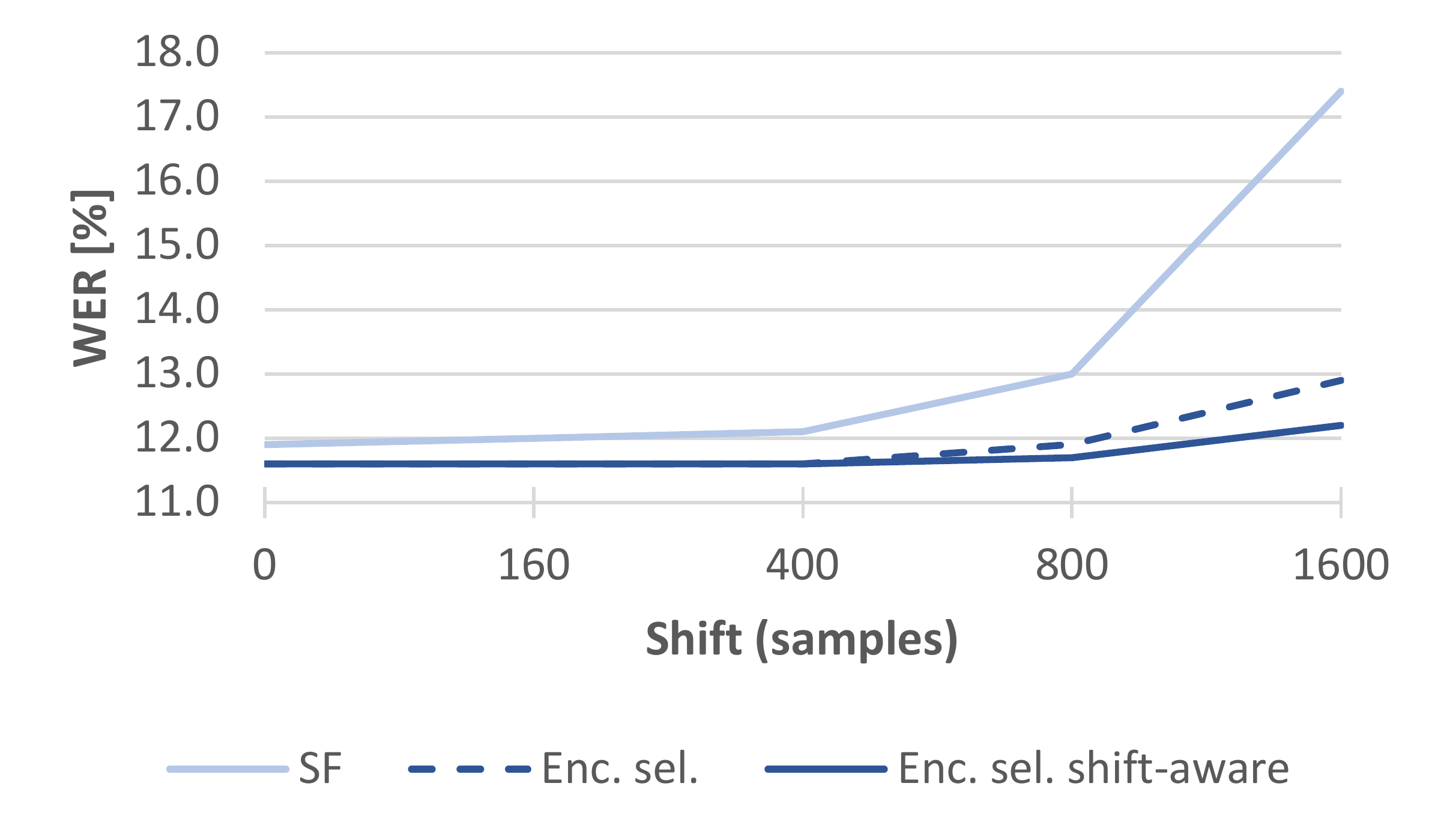}
    \caption{Performance of the Conformer Transducer using a single-encoder (SF on CT and FT channels) or dual-encoder architecture with encoder selection, for randomly shifted CT and FT inputs.}
    \label{fig:shifts}
\end{figure}

In both the SF on concatenated CT and FT input and the soft encoder selection, different input streams are combined frame-wise; thus, these methods can be affected by synchronization issues similar to those encountered in distributed microphone arrays \cite{Araki2018-MRW,HaebUmbach2021-FAS}.
To investigate the extent to which time offsets between FT and CT input can affect the performance, we artificially corrupted our test data (which is well synchronized) by shifting the CT vs.\ the FT input to simulate random offsets.
We evaluated SF and soft encoder selection for various maximum shifts up to 100\,ms (1600 samples) in both directions.
The results are shown in \figurename~\ref{fig:shifts}.
It can be seen that encoder selection is more robust against shifts than SF, which suffers from 46\,\% relative WER increase for a maximum shift of 1600 samples.
Yet, it still underperforms (i.e.\ it is worse than the single-encoder baselines) for shifts larger than 800 samples.

This led us to investigate {\em shift-aware training}, where we included random shifts during fine-tuning of the dual-encoder model.
In a proof-of-concept experiment, we set the maximum shift in training to 800 samples.
In this case, there is no significant degradation for shifts up to 800 samples in testing, and a graceful degradation for longer shifts not seen in training.
This kind of robustness is promising as exact synchronization can sometimes be hard to achieve.
However, more research is needed to fully assess the potential of data augmentation to cope with synchronization issues, also for the cases of (hard) encoder selection and sampling frequency mismatch \cite{Araki2018-MRW}.

\subsection{Discussion}

In the following, we would like to provide some insight into the rationale behind our encoder selection architecture, compared to alternative approaches for model combination.


\noindent \textbf{Model combination in the output space:} We compared our method to combining the ASR hypotheses of the single-encoder CT and FT systems via ROVER \cite{Fiscus1997-APS}. For the LAS architecture, we obtained 15.9\,\% WER, compared to 14.3\,\% with the proposed encoder selection architecture. The advantage of our approach is that it does not require time alignments or word-level confidences, which are not always easy to obtain from E2E systems.


\noindent \textbf{Encoder selection vs.\ stacking of encoder outputs:} We also experimented with an architecture that simply stacks the output of the CT and FT encoders. However, we could not achieve competitive performance in our scenario (the best WER was 12.6\,\%, compared to 11.6\,\% with the proposed method for the Conformer RNN-T). The reason is that the architecture with stacking requires training a new joint network layer (or decoder layer in case of the LAS architecture) from scratch. In our experiments, the fine-tuning started from a very high loss value and it only converged when freezing part of the model. In contrast, the proposed method with encoder selection can be trained easily with little data, since even a random encoder selection network provides reasonable ASR accuracy if the rest of the components is well initialized.
This is especially useful considering the effort to record parallel CT and FT data from a real usage scenario.

\section{Conclusion}

In this paper, we introduced a multi-encoder architecture with encoder selection to jointly model both CT and FT speech.
Our approach can be applied to various kinds of ASR systems, including the attention-based and RNN-T architectures investigated in this paper, but also the encoder-only Connectionist Temporal Classification (CTC) models and the classical hybrid system.
While hard selection can already outperform the dedicated CT and FT systems, the best performing method is the soft encoder selection, which relies on framewise combination of the encoder output.
To combat the sensitivity of this approach to the synchronization between CT and FT audio streams, we proposed shift-aware training.

In future work, we will 
extend our work to online streaming ASR, which can be supported by the framewise encoder selection approach.
We will also look into multi-task learning of the ASR and speaker role classification tasks to improve the performance of hard encoder selection. 
Finally, we will evaluate our architecture with several FT encoders using different types of beamforming.




\bibliographystyle{IEEEbib}
\bibliography{strings,refs}

\begin{thebibliography}{10}

\bibitem{VanCompernolle1990-SRI}
Dirk Van~Compernolle, Weiye Ma, Fei Xie, and Marc Van~Diest,
\newblock ``Speech recognition in noisy environments with the aid of microphone
  arrays,''
\newblock {\em Speech Communication}, vol. 9, no. 5, pp. 433--442, 1990.

\bibitem{Yoshioka2012-MMU}
Takuya Yoshioka, Armin Sehr, Marc Delcroix, Keisuke Kinoshita, Roland Maas,
  Tomohiro Nakatani, and Walter Kellermann,
\newblock ``Making machines understand us in reverberant rooms: Robustness
  against reverberation for automatic speech recognition,''
\newblock {\em IEEE Signal Processing Magazine}, vol. 29, no. 6, pp. 114--126,
  2012.

\bibitem{Li2014-AOO}
Jinyu Li, Li~Deng, Yifan Gong, and Reinhold Haeb-Umbach,
\newblock ``An overview of noise-robust automatic speech recognition,''
\newblock {\em IEEE/ACM Transactions on Audio, Speech, and Language
  Processing}, vol. 22, no. 4, pp. 745--777, 2014.

\bibitem{HaebUmbach2021-FAS}
Reinhold Haeb-Umbach, Jahn Heymann, Lukas Drude, Shinji Watanabe, Marc
  Delcroix, and Tomohiro Nakatani,
\newblock ``Far-field automatic speech recognition,''
\newblock {\em Proc.\ of the IEEE}, vol. 109, no. 2, pp. 124--148, 2021.

\bibitem{Graves2012-STW}
Alex Graves,
\newblock ``Sequence transduction with recurrent neural networks,''
\newblock in {\em Proc.\ of ICML Workshop on Representation Learning},
  Edinburgh, UK, 2012, PMLR.

\bibitem{Chan2016-LAA}
William Chan, Navdeep Jaitly, Quoc Le, and Oriol Vinyals,
\newblock ``Listen, attend and spell: A neural network for large vocabulary
  conversational speech recognition,''
\newblock in {\em Proc.\ of ICASSP}, Shanghai, China, 2016, pp. 4960--4964,
  IEEE.

\bibitem{Weninger2019-LAS}
Felix Weninger, Jes\'us Andr\'{e}s-Ferrer, Xinwei Li, and Puming Zhan,
\newblock ``Listen, {A}ttend, {S}pell and {A}dapt: Speaker adapted
  sequence-to-sequence {ASR},''
\newblock in {\em Proc.\ of INTERSPEECH}, Graz, Austria, 2019, pp. 3805--3809,
  ISCA.

\bibitem{Zeyer2019-ACO}
Albert Zeyer, Parnia Bahar, Kazuki Irie, Ralf Schlüter, and Hermann Ney,
\newblock ``A comparison of {T}ransformer and {LSTM} encoder decoder models for
  {ASR},''
\newblock in {\em Proc.\ of ASRU}, Sentosa, Singapore, 2019, pp. 8--15, IEEE.

\bibitem{Zhang2020-TTA}
Qian Zhang, Han Lu, Hasim Sak, Anshuman Tripathi, Erik McDermott, et~al.,
\newblock ``{Transformer Transducer}: A streamable speech recognition model
  with {T}ransformer encoders and {RNN\nobreakdash-T} loss,''
\newblock in {\em Proc.\ of ICASSP}, Barcelona, Spain, 2020, pp. 7829--7833,
  IEEE.

\bibitem{Gulati2020-CCA}
Anmol Gulati, James Qin, Chung-Cheng Chiu, Niki Parmar, Yu~Zhang, Jiahui Yu,
  Wei Han, Shibo Wang, Zhengdong Zhang, Yonghui Wu, et~al.,
\newblock ``{Conformer: Convolution-augmented {T}ransformer for speech
  recognition},''
\newblock in {\em Proc.\ of INTERSPEECH}, Shanghai, China, 2020, pp.
  5036--5040, ISCA.

\bibitem{Tuske2020-SHA}
Zolt{\'a}n T{\"u}ske, George Saon, Kartik Audhkhasi, and Brian Kingsbury,
\newblock ``Single headed attention based sequence-to-sequence model for
  state-of-the-art results on {Switchboard},''
\newblock in {\em Proc.\ of INTERSPEECH}, Shanghai, China, 2020, pp. 551--555,
  ISCA.

\bibitem{Sainath16-RTC}
Tara~N Sainath, Arun Narayanan, Ron~J Weiss, Ehsan Variani, Kevin~W Wilson,
  Michiel Bacchiani, and Izhak Shafran,
\newblock ``Reducing the computational complexity of multimicrophone acoustic
  models with integrated feature extraction,''
\newblock in {\em Proc.\ of INTERSPEECH}, San Francisco, CA, 2016, pp.
  1971--1975, IEEE.

\bibitem{Erdogan2016-IMB}
Hakan Erdogan, John~R Hershey, Shinji Watanabe, Michael~I Mandel, and Jonathan
  Le~Roux,
\newblock ``Improved {MVDR} beamforming using single-channel mask prediction
  networks,''
\newblock in {\em Proc.\ of INTERSPEECH}, San Francisco, CA, 2016, pp.
  1981--1985, ISCA.

\bibitem{Li2017-AMF}
Bo~Li, Tara~N Sainath, Arun Narayanan, Joe Caroselli, Michiel Bacchiani, Ananya
  Misra, Izhak Shafran, Hasim Sak, Golan Pundak, Kean~K Chin, et~al.,
\newblock ``Acoustic modeling for {Google Home},''
\newblock in {\em Proc.\ of INTERSPEECH}, Stockholm, Sweden, 2017, pp.
  399--403, ISCA.

\bibitem{Ochiai2017-UAF}
Tsubasa Ochiai, Shinji Watanabe, Takaaki Hori, John~R Hershey, and Xiong Xiao,
\newblock ``Unified architecture for multichannel end-to-end speech recognition
  with neural beamforming,''
\newblock {\em IEEE Journal of Selected Topics in Signal Processing}, vol. 11,
  no. 8, pp. 1274--1288, 2017.

\bibitem{Chang2020-EMS}
Xuankai Chang, Wangyou Zhang, Yanmin Qian, Jonathan~Le Roux, and Shinji
  Watanabe,
\newblock ``End-to-end multi-speaker speech recognition with {T}ransformer,''
\newblock in {\em Proc.\ of ICASSP}, Shanghai, China, 2020, pp. 6134--6138,
  IEEE.

\bibitem{Araki2018-MRW}
Shoko Araki, Nobutaka Ono, Keisuke Kinoshita, and Marc Delcroix,
\newblock ``Meeting recognition with asynchronous distributed microphone array
  using block-wise refinement of mask-based {MVDR} beamformer,''
\newblock in {\em Proc.\ of ICASSP}, Calgary, Canada, 2018, pp. 5694--5698,
  IEEE.

\bibitem{Yoshioka2019-MTU}
Takuya Yoshioka, Dimitrios Dimitriadis, Andreas Stolcke, William Hinthorn, Zhuo
  Chen, Michael Zeng, and Xuedong Huang,
\newblock ``Meeting transcription using asynchronous distant microphones,''
\newblock in {\em Proc.\ of INTERSPEECH}, Graz, Austria, 2019, pp. 2968--2972,
  ISCA.

\bibitem{Wang2020-EEM}
Xiaofei Wang, Naoyuki Kanda, Yashesh Gaur, Zhuo Chen, Zhong Meng, and Takuya
  Yoshioka,
\newblock ``Exploring end-to-end multi-channel {ASR} with bias information for
  meeting transcription,''
\newblock in {\em Proc.\ of IEEE Spoken Language Technology Workshop (SLT)},
  Shenzhen, China, 2021, pp. 833--840, IEEE.

\bibitem{Tang2018-ASO}
Hao Tang, Wei-Ning Hsu, Fran{\c{c}}ois Grondin, and James Glass,
\newblock ``A study of enhancement, augmentation, and autoencoder methods for
  domain adaptation in distant speech recognition,''
\newblock in {\em Proc.\ of INTERSPEECH}, Hyderabad, India, 2018, pp.
  2928--2932, ISCA.

\bibitem{Lippmann1987-MST}
Richard Lippmann, Edward Martin, and D~Paul,
\newblock ``Multi-style training for robust isolated-word speech recognition,''
\newblock in {\em Proc.\ of ICASSP}, Dallas, TX, 1987, vol.~12, pp. 705--708,
  IEEE.

\bibitem{Misra2003-NEB}
Hemant Misra, Herv{\'e} Bourlard, and Vivek Tyagi,
\newblock ``New entropy based combination rules in {HMM/ANN} multi-stream
  {ASR},''
\newblock in {\em Proc.\ of ICASSP}, Hong Kong, China, 2003, vol.~2, pp.
  741--744, IEEE.

\bibitem{Chen17-MTI}
Pin-Jung Chen, I-Hung Hsu, Yi-Yao Huang, and Hung-Yi Lee,
\newblock ``Mitigating the impact of speech recognition errors on chatbot using
  sequence-to-sequence model,''
\newblock in {\em Proc.\ of ASRU}, Okinawa, Japan, 2017, pp. 497--503.

\bibitem{Macoskey2021-BNA}
Jon Macoskey, Grant~P. Strimel, and Ariya Rastrow,
\newblock ``Bifocal neural {ASR}: Exploiting keyword spotting for inference
  optimization,''
\newblock in {\em Proc.\ of ICASSP}, Toronto, Canada, 2021, pp. 5999--6003,
  IEEE.

\bibitem{Narayanan2021-CEF}
Arun Narayanan, Tara~N. Sainath, Ruoming Pang, Jiahui Yu, Chung-Cheng Chiu,
  Rohit Prabhavalkar, Ehsan Variani, and Trevor Strohman,
\newblock ``Cascaded encoders for unifying streaming and non-streaming {ASR},''
\newblock in {\em Proc.\ of ICASSP}, Toronto, Canada, 2021, pp. 5629--5633,
  IEEE.

\bibitem{Braun2018-MCA}
Stefan Braun, Daniel Neil, Jithendar Anumula, Enea Ceolini, and Shih-Chii Liu,
\newblock ``Multi-channel attention for end-to-end speech recognition,''
\newblock in {\em Proc.\ of INTERSPEECH}, Hyderabad, India, 2018, pp. 17--21,
  ISCA.

\bibitem{Chang2021-EMT}
Feng-Ju Chang, Martin Radfar, Athanasios Mouchtaris, Brian King, and Siegfried
  Kunzmann,
\newblock ``End-to-end multi-channel {T}ransformer for speech recognition,''
\newblock in {\em Proc.\ of ICASSP}, Toronto, Canada, 2021, pp. 5884--5888,
  IEEE.

\bibitem{Wang2019-SAB}
Xiaofei Wang, Ruizhi Li, Sri~Harish Mallidi, Takaaki Hori, Shinji Watanabe, and
  Hynek Hermansky,
\newblock ``Stream attention-based multi-array end-to-end speech recognition,''
\newblock in {\em Proc.\ of ICASSP}, Brighton, UK, 2019, pp. 7105--7109, IEEE.

\bibitem{Li2020-MSE}
Ruizhi Li, Xiaofei Wang, Sri~Harish Mallidi, Shinji Watanabe, Takaaki Hori, and
  Hynek Hermansky,
\newblock ``Multi-stream end-to-end speech recognition,''
\newblock {\em IEEE/ACM Transactions on Audio, Speech, and Language
  Processing}, vol. 28, pp. 646--655, 2020.

\bibitem{Li2021-TAA}
Ruizhi Li, Gregory Sell, and Hynek Hermansky,
\newblock ``Two-stage augmentation and adaptive {CTC} fusion for improved
  robustness of multi-stream end-to-end {ASR},''
\newblock in {\em Proc.\ of IEEE Spoken Language Technology Workshop (SLT)},
  Shenzhen, China, 2021, pp. 229--235, IEEE.

\bibitem{Li20-APT}
Ruizhi Li, Gregory Sell, Xiaofei Wang, Shinji Watanabe, and Hynek Hermansky,
\newblock ``A practical two-stage training strategy for multi-stream end-to-end
  speech recognition,''
\newblock in {\em Proc.\ of ICASSP}, Barcelona, Spain, 2020, pp. 7014–--7018,
  IEEE.

\bibitem{Bahdanau2015-NMT}
Dzmitry Bahdanau, Kyunghyun Cho, and Yoshua Bengio,
\newblock ``Neural machine translation by jointly learning to align and
  translate,''
\newblock in {\em Proc.\ of ICLR}, San Diego, CA, 2015, open publishing.

\bibitem{Norouzian2019-EAM}
Atta Norouzian, Bogdan Mazoure, Dermot Connolly, and Daniel Willett,
\newblock ``Exploring attention mechanism for acoustic-based classification of
  speech utterances into system-directed and non-system-directed,''
\newblock in {\em Proc.\ of ICASSP}, Brighton, UK, 2019, pp. 7310--7314, IEEE.

\bibitem{Neumann2017-ACN}
Michael Neumann and Ngoc~Thang Vu,
\newblock ``Attentive convolutional neural network based speech emotion
  recognition: A study on the impact of input features, signal length, and
  acted speech,''
\newblock in {\em Proc.\ of INTERSPEECH}, Stockholm, Sweden, 2017, pp.
  1263--1267, ISCA.

\bibitem{Wu2019-FDM}
Minhua Wu, Kenichi Kumatani, Shiva Sundaram, Nikko Str{\"o}m, and Bj{\"o}rn
  Hoffmeister,
\newblock ``Frequency domain multi-channel acoustic modeling for distant speech
  recognition,''
\newblock in {\em Proc.\ of ICASSP}, Brighton, UK, 2019, pp. 6640--6644, IEEE.

\bibitem{Kumatani2019-MGS}
Kenichi Kumatani, Minhua Wu, Shiva Sundaram, Nikko Str{\"o}m, and Bj{\"o}rn
  Hoffmeister,
\newblock ``Multi-geometry spatial acoustic modeling for distant speech
  recognition,''
\newblock in {\em Proc.\ of ICASSP}, Brighton, UK, 2019, pp. 6635--6639, IEEE.

\bibitem{Weninger2020-SSL}
Felix Weninger, Franco Mana, Roberto Gemello, Jes\'us Andr\'{e}s-Ferrer, and
  Puming Zhan,
\newblock ``Semi-supervised learning with data augmentation for end-to-end
  {ASR},''
\newblock in {\em Proc.\ of INTERSPEECH}, Shanghai, China, 2020, pp.
  2802--2806, ISCA.

\bibitem{Park2019-SAAa}
Daniel~S. Park, William Chan, Yu~Zhang, Chung-Cheng Chiu, Barret Zoph, Ekin~D.
  Cubuk, and Quoc~V. Le,
\newblock ``{SpecAugment}: A simple data augmentation method for automatic
  speech recognition,''
\newblock in {\em Proc.\ of INTERSPEECH}, Graz, Austria, 2019, pp. 2613--2617,
  ISCA.

\bibitem{Gillick1989-SSI}
Laurence Gillick and Stephen~J Cox,
\newblock ``Some statistical issues in the comparison of speech recognition
  algorithms,''
\newblock in {\em Proc.\ of ICASSP}, Glasgow, UK, 1989, vol.~1, pp. 532--535,
  IEEE.

\bibitem{Fiscus1997-APS}
Jonathan~G. Fiscus,
\newblock ``{A post-processing system to yield reduced word error rates:
  Recognizer Output Voting Error Reduction (ROVER)},''
\newblock in {\em Proc.\ of ASRU}, Santa Barbara, CA, 1997, pp. 347--354, IEEE.

\end{thebibliography}

\end{document}